\documentstyle[aps,epsfig,rotating,color]{revtex}

\def\beq{\begin{equation}}
\def\eeq{\end{equation}}

\def\reff#1{(\ref{#1})}

\def\vekt#1{\bbox{#1}}

\def\vektr{\vekt{r}}

\def\vekte{\vekt{e}}
\def\vektE{\vekt{E}}

\def\vektnabla{\vekt{\nabla}}

\def\operator#1{\mbox{\sf #1}}

\def\hamop{{\operator{H}}}

\def\halb{\frac{1}{2}}

\def\Edach{\hat{E}}
\def\Ehat{\hat{E}}

\def\energy{{\cal{E}}}
\def\Ptrans{{\cal{P}}}

\def\pabl#1#2{\frac{\partial #1}{\partial #2}}

\def\imagi{\mbox{\rm i}}
\def\diff{\,\mbox{\rm d}}

\def\bild#1#2{\epsfig{file={#1},width=#2}}

\begin{document}
%\draft

\title{A numerical ab initio study of harmonic generation from a ring-shaped model
  molecule in laser fields}
\author{D.~Bauer and  F.~Ceccherini}
\address{Theoretical Quantum Electronics (TQE), Darmstadt University of
  Technology,\\ Hochschulstr.\ 4A, D-64289 Darmstadt, Germany}
\date{\today}

\maketitle

\begin{abstract}
When a laser pulse impinges on a molecule which is invariant
under certain symmetry operations selection rules for harmonic
generation (HG) arise. 
In other words: symmetry
controls which channels are open for the deposition and emission of laser energy---with
the possible application of filtering or amplification. We review the
derivation of HG selection rules and study numerically the
interaction of laser pulses with an effectively one-dimensional ring-shaped
model molecule. 
The harmonic yields obtained from that model and their dependence on laser
frequency and intensity are discussed.
In a real experiment obvious 
candidates for such molecules are benzene, other aromatic compounds, or even
nanotubes. 
\end{abstract}


%\vspace{-3.5cm}

%\hspace{-1cm}
%\begin{rotate}{30}{
%\color{red}
%\fbox{\color{red} \Huge\bf D R A F T}}
%\end{rotate}

%\vspace{2cm}

%\quad\hspace{13cm}\parbox{1cm}{
%\begin{rotate}{35}{
%\color{red}
%\fbox{\color{red} \LARGE\bf to be completed }}
%\end{rotate}}

\section{Introduction}
Harmonic generation (HG) from atoms (L'Huillier \& Balcou 1993), molecules
(Liang et al.\ 1994), clusters (Donelly et al.\ 1996), and solids (von der
Linde et al.\ 1995) as a short-wavelength source is of great practical relevance. In recent years
a huge amount of publications were devoted to ``harmonic engineering,'' under
which we subsume either the study of phase matching during the propagation of
the emitted light through gaseous media (Gaarde et al.\ 1998), generation of
attosecond pulses using varying ellipticity (Antoine et al.\ 1997), multi-color
studies (Milosevic et al.\ 2000), or HG from thin crystals (Faisal \& Kaminski
1996 and 1997). Experimental results on HG from cyclic organic molecules were
also reported in the literature (Hay et al.\ 2000a and 2000b). However, the
authors made no attempt to verify theoretically predicted selection rules.
In the well established physical picture of HG one assumes that
an electron tunnels out of the atom or ion, moves in the laser field and
eventually rescatters with its parent (or other) ion where it might
recombine---leading to an emission of a photon with several times the
fundamental frequency (Becker et al.\ 1997). Assuming such a
viewpoint one can easily explain prominent features of HG spectra 
such as the famous cut-off at $I_p + 3.17 U_p$ in the single atom case 
(where $I_p$ is the ionization energy and
$U_p$ is the ponderomotive energy, i.e., the cycle-averaged quiver energy of
the electron in the laser field). For a general overview on harmonic
generation in laser fields see the recent review by Sali\`eres et al.\ (1999).

Fortunately, to derive the selection rules for harmonic emission only symmetry
considerations are necessary (Alon et al.\ 1998). We now briefly summarize this
approach. Let us consider a Hamiltonian which is periodic in time, i.e.,
$\hamop(t)=\hamop(t+\tau)$. Such a Hamiltonian might describe an electron in a
{\em long} laser pulse where the pulse envelope is sufficiently adiabatic. The
time-dependent Schr\"odinger equation (TDSE) reads (we use atomic units
$\hbar=e=m=1$ throughout, if not noted otherwise)
\beq \left[ \hamop(t) - \imagi \pabl{}{t} \right] \Psi_\energy
(\vektr,t) = 0. \eeq
According to the Floquet theorem (see, e.g., Faisal 1987) we can write
$\Psi_\energy(\vektr,t)=\psi_\energy(\vektr,t)\exp(-\imagi\energy t)$ with
$\psi_\energy(\vektr,t)=\psi_\energy(\vektr,t+\tau)$ leading to the Schr\"odinger equation 
\beq \hamop_F(t) \psi_\energy(\vektr,t) = \energy  \psi_\energy(\vektr,t), \qquad
\hamop_F(t)=\hamop(t)-\imagi\pabl{}{t} \eeq
with $\energy$ the so-called quasi energy and $\hamop_F$ the Floquet
Hamiltonian. 
Since $\psi_\energy(\vektr,t)$ is
periodic in time it might be expanded in a Fourier series
$\psi_\energy(\vektr,t)=\sum_n \exp[-\imagi n (\omega t+\delta)]
\Phi^{(n)}_\energy(\vektr)$. This Floquet approach is a well-known method in
multiphoton physics (Gavrila 1992) and a widely used numerical simulation technique
also (Potvliege 1998). In order to derive the HG selection rules we now assume (for
reasons which will become clear soon) that the
system is in a single and non-degenerate Floquet state $\psi_\energy$.
The HG spectra (HGS) peak no.\ $n$ is present only if the Fourier transformed
dipole moment $\mu(\vektr)$ does not vanish for the frequency $n\omega$,
\beq \int \diff t\, \exp(-\imagi n\omega t) \int\diff^3 r\,
\Psi^*_\energy(\vektr,t) \mu(\vektr) \Psi_\energy(\vektr,t) \neq 0.\label{fourier} \eeq 
Now let us suppose we worked out a symmetry operation $\Ptrans$ under which the
Floquet Hamiltonian is invariant, $\Ptrans \hamop_F(t)\Ptrans^{-1}=\hamop_F(t)$. 
It follows that (in the case of non-degenerated Floquet states
$\psi_\energy$, see assumption above) $\Ptrans\psi_\energy=
a \psi_\energy$ holds. Here $a$ is a phase factor, $\vert a\vert =1$.
For convenience, we rewrite \reff{fourier} as
\beq \langle\langle\Psi_\energy\vert\mu(\vektr)\exp(-\imagi n\omega
t)\vert\Psi_\energy\rangle\rangle =  \langle\langle\psi_\energy\vert\mu(\vektr)\exp(-\imagi  n\omega
t)\vert\psi_\energy\rangle\rangle  \neq 0\eeq
where the double brackets indicate spatial {\em and} temporal integration
[this is the so-called extended Hilbert space formalism, see, e.g., Sambe (1973)].
The HG selection rule can be derived from
\[  \langle\langle\psi_\energy\vert\mu(\vektr)\exp(-\imagi  n\omega
t)\vert\psi_\energy\rangle\rangle =  \langle\langle\Ptrans\psi_\energy\vert\Ptrans\mu(\vektr)\exp(-\imagi  n\omega
t)\Ptrans^{-1}\vert\Ptrans\psi_\energy\rangle\rangle \]
\beq =\langle\langle\psi_\energy\vert\Ptrans\mu(\vektr)\exp(-\imagi  n\omega
t)\Ptrans^{-1}\vert\psi_\energy\rangle\rangle \label{step} \eeq
leading to
\beq \mu(\vektr)\exp(-\imagi  n\omega t) = \Ptrans\mu(\vektr)\exp(-\imagi  n\omega t)\Ptrans^{-1}. \label{selection} \eeq
The last step in \reff{step} is not possible if $\psi_\energy$ is not a pure
Floquet state. In the next Section we apply \reff{selection} to derive the
selection rule for HG from ring-shaped molecules. Here, for the sake of
illustration, we rederive the selection rule for a single atom with
spherically symmetric potential $V(r)$ and linearly (in $x$-direction) polarized laser field
$\vektE(t)=\Edach\vekte_x\sin\omega t$. The Floquet Hamiltonian $\hamop_F(t)=-\halb
\vektnabla^2 + V(r) + \Edach x \sin\omega t - \imagi\partial_t$ is
invariant under the transformation $x\to -x$ and $t\to t+\pi/\omega$.
If we look for harmonics polarized in the $x$-direction we have
$\mu(\vektr)=x$ and from \reff{selection} $x \exp(-\imagi  n\omega t) = -x \exp[-i
n\omega (t+\pi/\omega)]$ follows $\exp(-\imagi n \pi)=-1$.
Therefore only odd harmonics are generated.

The paper is organized as follows: in Section \ref{model} we introduce our
model and the selection rule which holds in its case. In Section \ref{results}
we present and discuss our numerical results. Finally, in Section \ref{summ}
we give a summary and an outlook.

\section{A simple one-dimensional model for ring-like molecules} \label{model}
The time-dependent Schr\"odinger equation (TDSE) for a single electron in a laser
field $\vektE(t)$ and under the influence of an ionic potential $V(\vektr)$ 
reads in dipole approximation and length gauge
\beq \imagi \pabl{}{t} \Psi(\vektr,t) = \left( -\halb \vektnabla^2 + V(\vektr)
  + \vektE(t )\cdot \vektr \right) \Psi(\vektr,t) . \label{tdse.i} \eeq
The dipole approximation is excellent since in all the cases studied in this
paper the wavelength of the laser light is much greater than the size of the
molecule.

If we force the electron to move along a ring of radius $\rho$ in the
$xy$-plane $V(\vektr)$ becomes $V(\varphi)$ where $\varphi$ is the usual polar
angle. With an electric field of the form $  \vektE(t)=\Ehat(t) \left[ \xi \cos(\omega t) \vekte_x + \sqrt{1-\xi^2}
  \sin(\omega t)\vekte_y \right] $, 
where $\Ehat(t)$ is a slowly varying envelope, $\omega$ is the laser frequency,
and $\xi$ is the ellipticity
parameter, 
the TDSE \reff{tdse.i}
becomes effectively one-dimensional (1D) in space and reads
\beq \imagi \pabl{}{t} \Psi(\varphi,t) = \left( -\frac{1}{2\rho^2}
  \pabl{^2}{\varphi^2} + V(\varphi) + \Ehat(t)\rho \left[ \xi\cos(\omega
    t)\cos\varphi + \sqrt{1-\xi^2} \sin(\omega t)\sin\varphi\right] \right)
\Psi(\varphi,t). \label{tdse.ii} \eeq
In case of circularly polarized light ($\xi=1/\sqrt{2}$) this simplifies to
\beq \imagi \pabl{}{t} \Psi(\varphi,t) = \left( -\frac{1}{2\rho^2}
  \pabl{^2}{\varphi^2} + V(\varphi) + \frac{\Ehat(t)\rho}{\sqrt{2}} \cos(\varphi-\omega t) \right)
\Psi(\varphi,t). \label{tdse.iii} \eeq

We now assume that the potential $V(\varphi)$ has an $N$-fold rotational
symmetry, $V(\varphi+2\pi/N)=V(\varphi)$. 
Then, with the help of \reff{selection}, we can easily derive the selection rule for
HG in the system described by the TDSE \reff{tdse.iii}. The transformation
($\varphi\to\varphi+2\pi/N$, $t\to t + 2\pi/N\omega$) leaves the corresponding
Floquet Hamiltonian invariant. For, e.g.,  anti-clockwise polarized emission
$\mu(\vektr)=\rho\exp(\imagi\varphi)$ holds, and
from \reff{selection} we have
\[ \rho \exp(\imagi\varphi)\exp(-\imagi n \omega t) = \rho
\exp[\imagi(\varphi+2\pi/N)]\exp[-\imagi n \omega (t+2\pi/N\omega)] \]
leading to $n=Nk+1$, $k=1,2,3 ...$.
For the clockwise emission one finds accordingly $n=Nk-1$. Thus we expect
pairs of HG peaks at $kN\pm 1$ (Alon et al.\ 1998).

The TDSE \reff{tdse.ii} or \reff{tdse.iii} can be easily solved ab
  initio on a PC. We did this by propagating the wavefunction in time with a
Crank-Nicholson approximant to the propagator $U(t+\Delta
t,t)=\exp[-\imagi\Delta t \hamop(t+\Delta t/2)]$ where $\hamop(t)$ is the
explicitly time-dependent Hamiltonian corresponding to the TDSE
\reff{tdse.ii}. Our algorithm is fourth order in the grid spacing
$\Delta\varphi$ and second order in the time step $\Delta t$.
The boundary condition is $\Psi(0,t)=\Psi(2\pi,t)$ for all times $t$. 

\section{Numerical results and discussion} \label{results}
We now present results from single active electron (SAE)-runs with $\rho=2.64$
(bond length and radius of benzene C$_6$H$_6$) and an effective model potential
\beq V(\varphi) = -\frac{V_0}{2} [\cos(N\varphi) +1] \label{pot} \eeq
with $N=6$ and $V_0=0.6405$. This leads to an electronic ground state energy
$\energy_0=-0.34$ which is the experimental ionization potential for
removing the first electron in benzene (see, e.g., Talebpour et al.\ 1998 and
2000). 
Note, that in our simple model
we have no continuum but discrete states only. 
The first six excited states are located at $-0.27$
(two-fold degenerated), $-0.07$ (two-fold degenerated), $0.16$
(non-degenerated), $0.48$ (non-degenerated), $0.85$
(two-fold degenerated), $1.48$ (two-fold degenerated).
The energy levels of our model resemble, apart from an
overall downshift and the removal of degeneracies of certain states, those of
the isoperimetric model where $V_0=0$, the energy levels are given by 
$\energy_m=m^2/2\rho^2$, $m=0,1,2,\ldots$
with the states $m\neq 0$ two-fold degenerated. Therefore, the energy level
spacing, and thus typical electronic transitions, are different from those in
real benzene. However, it is not our goal to present quantitatively correct
results for laser benzene-interaction in this paper but we rather want to
demonstrate some of the underlying principles of HG from ring-shaped molecules in general. 

In Fig.~\ref{circandlin} we present HGS for a $q=240$ 
cycle pulse of the shape $\Ehat(t)=\Ehat \sin^2\left(\omega t/2
    q\right)$ with an electric field amplitude
$\Edach=0.5$~a.u.\ and frequency $\omega=0.18$. In  Fig.~\ref{circandlin}a the
result for linear polarization $\xi=1$ is shown, in Fig.~\ref{circandlin}b
the result for circular polarization $\xi=1/\sqrt{2}$. To obtain those plots
we evaluated the Fourier transformed dipole in $x$-direction, i.e.,
$\langle\langle\Psi(t)\vert\rho\cos(\varphi)\exp(-\imagi n\omega t)
\vert\Psi(t)\rangle\rangle$. In the $\xi=1$-case the dipole with respect to
$y$ is clearly zero while in the circular case there is simply a phase shift
of $\pi/2$ with respect to the dipole in $x$. As expected, in the linearly
polarized field all odd harmonics are emitted whereas in the circular case
only the harmonics $6k\pm 1$, $k=1,2,\ldots$ are visible. Other emission lines
are many orders of magnitude weaker. Those lines can  form band-like structures which
are interesting in itself. However, in this paper we focus only on the laser harmonics.
They dominate the HGS, at least as long as no resonances are hit. 

In Fig.~\ref{difffrequ} the emitted yield of the fundamental and the first
four harmonics (5th, 7th, 11th, 13th) in a circularly polarized laser pulse 
are presented as a function of the laser
frequency. The pulse length
$T=8378$~a.u.\ (corresponding to $\approx 200$~fs) and peak field
strength $\Edach=0.2$~a.u.\
(corresponding to $1.4\times 10^{15}$~W/cm$^2$) were held fixed. The frequency is
plotted in units of the smallest level spacing of the model molecule, i.e.,
the energy gap between first excited and ground state, which is in our case
$\Omega=0.34-0.27=0.07$. For laser frequencies $\omega < \Omega$ we do not
find the 7th or higher harmonics. The 11th and 13th harmonic show an overall
decrease with increasing laser frequency whereas the fundamental, the 5th and
the 7th stay relatively constant in intensity. 
For
frequencies $\omega<2.5\Omega$ there is a complicated dependency of the
harmonic yield on $\omega$. 
All harmonics
show a local maximum around $1.3\Omega$. However, at that frequency one
apparently hits a resonance since the harmonic peaks become broad and show a 
substructure (see left inlay in Fig.~\ref{difffrequ}). In the interval
$2<\omega/\Omega<2.5$ the fundamental drops whereas the 7th harmonic increases
in strength. Note that the 7th harmonic is anti-clockwise polarized, like the
incident laser field, whereas the 5th is polarized in the opposite direction.
For frequencies $\omega>2.5\Omega$ the behavior becomes
more smooth apart from another resonance near $3.8\Omega$.
In general, the HGS look clean for sufficiently high frequencies and far away from resonances
(like in the right inlay of Fig.~\ref{difffrequ}). A rich substructure
near resonances is visible in the HGS (cf.\ inlay for $\omega=2\Omega$). 

In Fig.~\ref{difffield} harmonic yields for the fixed frequency
  $\omega=2.8\Omega$ as a function of the  field amplitude $\Edach$ are shown. The higher
  harmonics (11th--19th) appear only for higher field strengths whereas
  fundamental, 5th and 7th are rather weakly dependent on the field strength. 
The anti-clockwise polarized harmonics (polarized like the incident laser
  light, drawn thick) tend to overtake the clockwise
  polarized ones (drawn thin) at higher laser intensities. However,
  $\Edach=0.6$ corresponds already to a laser intensity $1.3\times
  10^{16}$~W/cm$^2$ where a real benzene molecule would probably break.

It is interesting to study the scaling of the TDSE \reff{tdse.ii} with respect
to the size of the molecule. If one
scales the molecule radius like $\rho'=\alpha\rho$,
the TDSE  \reff{tdse.ii} remains invariant if 
$t'=\alpha^2 t$, $V'=V/\alpha^2$, $E'=E/\alpha^3$, 
$\omega'=\omega/\alpha^2$
is chosen. From this and our numerical result that laser frequencies $>\Omega$ are
preferable for clean HGS we learn that molecules bigger than benzene are more
promising candidates for HG with realistic laser frequencies [from Nd
($\omega=0.04$~a.u.) to KrF ($\omega=0.18$~a.u.)].

One might object that electron correlation could spoil the selection rule because in
reality it is not only a single electron which participates in the
dynamics. However, Alon et al.\ (1998) have proven  that this is not the
case. This is due to the fact that {\em (i)} the electron interaction part of the
Hamiltonian is still invariant under the transformation $\Ptrans$,
and {\em (ii)}   $\Ptrans$ commutes with the (anti-) symmetrization operator.
Even approximate theories or numerical techniques like time-dependent
Hartree-Fock or density functional theory do not spoil the selection rule
 since they involve only functionals which depend on scalar products of single
 particle orbitals,  all invariant under the transformation $\Ptrans$
 (Ceccherini \& Bauer 2001).

\section{Summary and outlook}\label{summ}
In this paper we demonstrated numerically HG from a model molecule with
discrete rotational symmetry which is subject to a circularly polarized laser
pulse. 
In particular the harmonic emission from an effectively 1D model with $N=6$
(i.e., a simple model for benzene) as a function of laser frequency and
intensity was discussed. It was found that for frequencies below the
characteristic level spacing $\Omega$ HG is strongly affected by
resonances. The situation relaxes for higher frequencies. For the efficient
generation of higher harmonics laser frequencies $>\Omega$ {\em and} rather strong
fields are necessary. In such fields real aromatic compounds probably ionize and
dissociate already.

Numerical studies for a more realistic, effectively 2D model molecule will be
presented elsewhere (Ceccherini \& Bauer 2001).

In order to obtain short wavelength radiation it is desirable to have either
$k$ or $N$ in the selection rule
$kN\pm 1$ as big as possible.  From our numerical results we infer that it is
probably hard to push efficiently towards high $k$ without destroying the
target molecule. For that reason,  in a real experiment nanotubes are  promising
candidates because $N$ can be of the order of 100 or more (Dresselhaus et al.\
1998), and, moreover, HG
should be even more efficient when the laser propagates through the tube. 
However, it remains the problem of the proper alignment of the laser
beam and the symmetry axis of the molecule. Crystals might be better
candidates in that respect.

\section*{Acknowledgement}
This work was supported by the Deutsche Forschungsgemeinschaft in the
framework of the SPP ``Wechselwirkung intensiver Laserfelder mit Materie.''

\pagebreak

\begin{center}
\large REFERENCES
\end{center}

\noindent  Alon O., Averbukh, V.\ \& Moiseyev, N.\ 1998 Phys.\ Rev.\ Lett.\
{\bf 80} 3743. \\
Antoine Ph., Milosevic, D.\ B.,  L'Huillier, A.,  Gaarde, M.\ B.,
Sali\`eres P.\ \& Lewenstein, M.\ 1997, Phys.\ Rev.\ A {\bf 56} 4960.  \\
Becker, W., Lohr, A.,  Kleber, M.\ \& Lewenstein, M.\ 1997, Phys.\ Rev.\ A
{\bf 56} 645, and references therein. \\
Ceccherini, F.\ \& Bauer, D.\ 2001, in preparation. \\
Donelly, T.\ D., Ditmire, T., Neuman, K., Perry, M.\ D.\ \& Falcone, R.\
W.\ 1996, Phys.\ Rev.\ Lett.\ {\bf 76} 2472.\\
Dresselhaus, M., Dresselhaus, G., Eklund, P.\ \& Saito, R.\ 1998, Physics
World {\bf 11} issue 1 article 9, {\tt
  http://physicsweb.org/article/world/11/1/9}\ .\\
Faisal, F.\ H.\ M.\ 1987 ``Theory of Multiphoton Processes,'' Plenum Press,
New York. \\
Faisal,  F.\ H.\ M.\ \& Kaminski, J.\ Z.\ 1996, Phys.\ Rev.\ A {\bf 54} R1769.\\
Faisal,  F.\ H.\ M.\ \& Kaminski, J.\ Z.\ 1997, Phys.\ Rev.\ A {\bf 56} 748.\\
Gaarde, M.\ B., Antoine, Ph., L'Huillier, A., Schafer, K.\ J.\ \& Kulander,
K.\ C.\ 1998,  Phys.\ Rev.\ A {\bf 57} 4553, and references therein.\\
Gavrila M.\ 1992, in: ``Atoms in Intense Laser Fields,'' ed.\ by Gavrila, M.,
Academic Press, New York, p.\ 435.\\
Hay, N., de Nalda, R., Halfmann, T., Mendham, K.\ J., Mason, M.\ B.,
Castillejo, M.\ and Marangos, J.\ P.\ 2000a, Phys.\ Rev.\ A {\bf 62}
041803(R).\\
Hay, N., Castillejo, M., de Nalda, R., Springate, E.,  Mendham, K.\ J.\ \&
Marangos, J.\ P.\ 2000b, Phys.\ Rev.\ A {\bf 62} 053810.\\
von der Linde, D., Engers, T., Jenke, G., Agostini, P., Grillon, G., Nibbering, E.,
Mysyrowicz, A.\ \& Antonetti A.\ 1995, Phys.\ Rev.\ A {\bf 52} R25. \\ 
L'Huillier, A.\ \& Balcou, Ph.\ 1993, Phys.\ Rev.\ Lett.\ {\bf 70} 774.
\\
Liang, Y., Augst, S., Chin, S.\ L., Beaudoin, Y.\ \& Chaker M.\ 1994, J.\ Phys.\
B: At.\ Mol.\ Opt.\ Phys.\ {\bf 27} 5119. \\
Milosevic, D.\ B., Becker, W.\ \& Kopold, R.\ 2000, Phys.\ Rev.\ A {\bf 61}
063403. \\
Potvliege, R.\ M.\ 1998, Comp.\ Phys.\ Comm.\ {\bf 114} 42. \\
Sali\`eres, P., L'Huillier, A., Antoine, Ph.\ \& Lewenstein, M.\ 1999, Adv.\
At.\ Mol.\ Opt.\ Phys.\ {\bf 41} 83.\\
Sambe H.\ 1973,  Phys.\ Rev.\ A {\bf 7} 2203. \\ 
Talebpour, A., Larochelle, S.\ \&  Chin, S.\ L.\ 1998, J.\
Phys.\ B: At.\ Mol.\ Opt.\ Phys.\ {\bf 31} 2769, and references therein. \\
Talebpour, A., Bandrauk, A.\ D., Vijayalakshmi, K.\ \& Chin, S.\ L.\ 2000, J.\
Phys.\ B: At.\ Mol.\ Opt.\ Phys.\ {\bf 33} 4615, and references therein. \\

\begin{figure}
\caption{\label{circandlin} HGS for a $240$ 
cycle $\sin^2$-shaped pulse with an electric field amplitude
$\Edach=0.5$~a.u.\ and frequency $\omega=0.18$. The polarization was (a)
linear ($\xi=1$) and (b) circular ($\xi=1/\sqrt{2}$). In (a) odd harmonics up
to $n=23$ dominate whereas in (b) the harmonics obey the selection rule
$6k\pm1$, $k=1,2,\ldots$, i.e., the 5th, 7th, 11th, 13th, 17th, 19th harmonics
are visible.}
\end{figure} 

\begin{figure}
\caption{\label{difffrequ} Fundamental and harmonic yield vs.\ laser frequency $\omega$
  in units of the smallest level spacing $\Omega=0.07$. Laser field intensity and
  pulse length was kept fixed. The anti-clockwise polarized fundamental, 7th,
  and 13th are plotted thick, the clockwise polarized 5th and 11th are drawn
  thin.
The inlays show spectra (harmonic yield vs.\ harmonic order) for three particular frequencies (indicated by
  arrows). 
See text for further discussion.}
\end{figure}

\begin{figure}
\caption{\label{difffield} Harmonic and fundamental yields for fixed frequency
  $\omega=2.8\Omega$ but different field amplitude $\Edach$. 
The inlays show spectra (harmonic yield vs.\ harmonic order) for three particular $\Edach$ (indicated by
  arrows). See text for further discussion.}
\end{figure} 

\pagebreak

\thispagestyle{empty}

\unitlength1cm
\begin{picture}(14,19)
\put(2,7){\bild{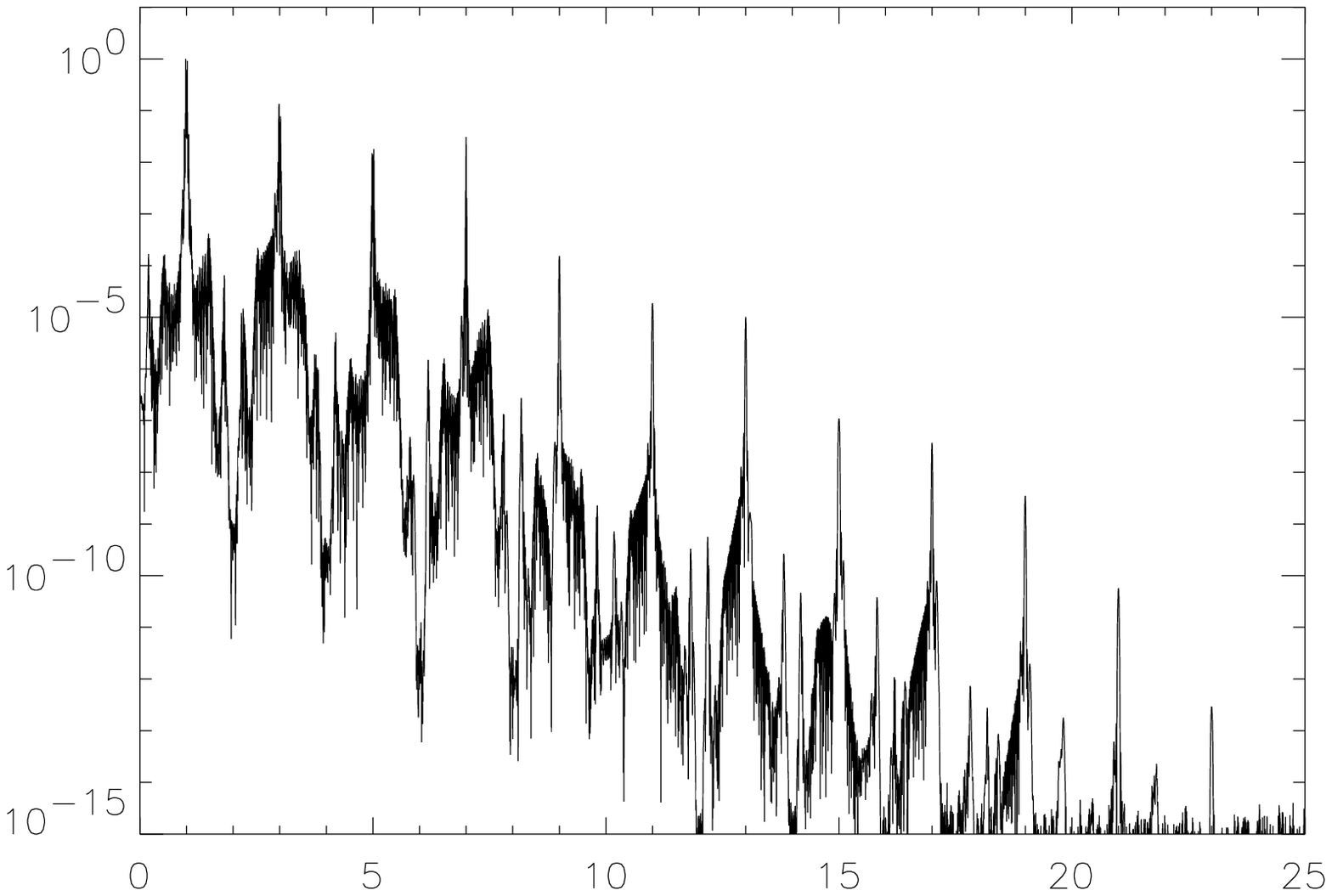}{11cm}}
\put(2,0){\bild{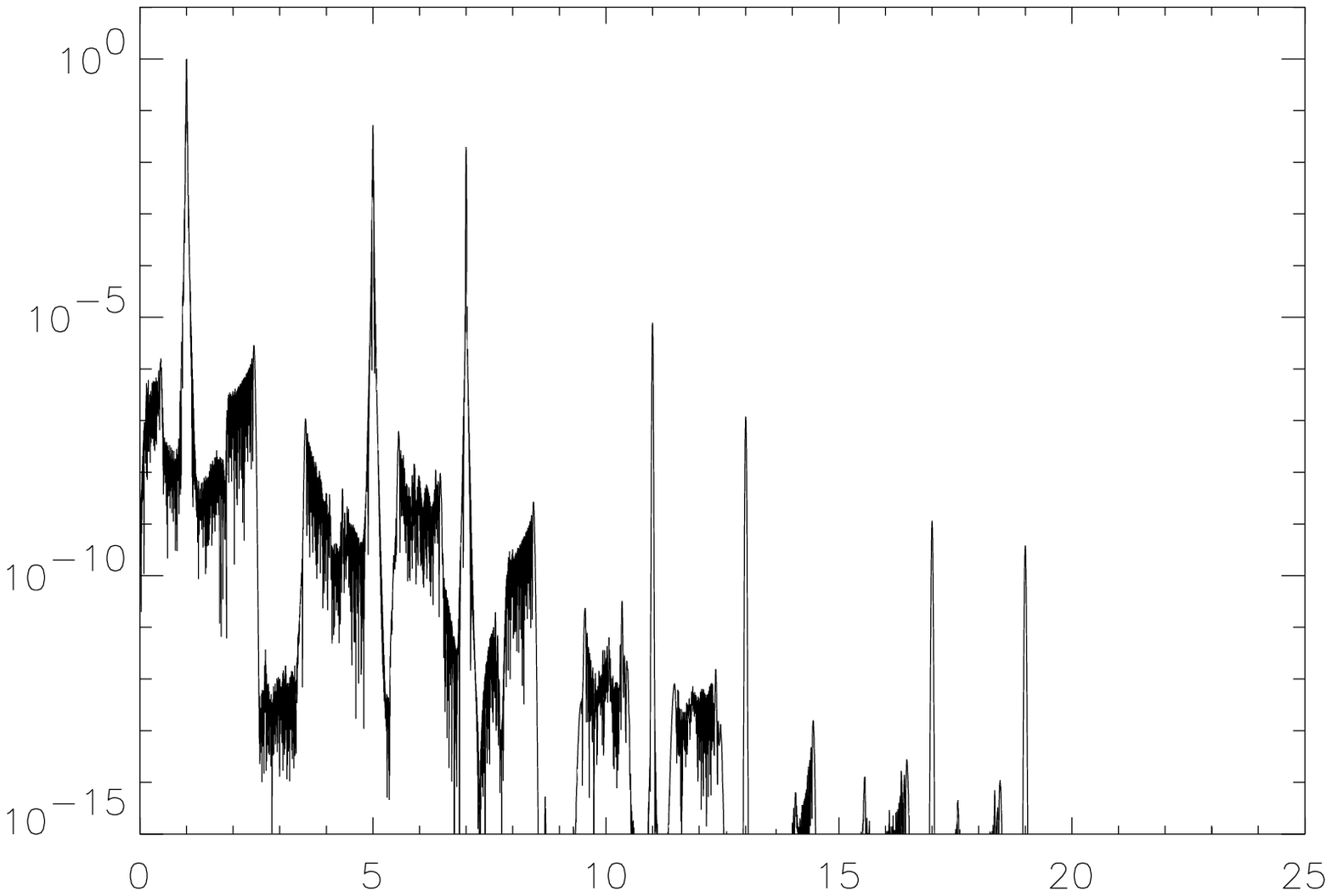}{11cm}}
\put(6.8,0){Harmonic order}
\put(1.8,5.5){\begin{sideways} Harmonic yield (arb.\ u.) \end{sideways}}
\put(10,13.5){(a) linear}
\put(10,6.5){(b) circular}
\end{picture}

\vfill

{\bf Fig. 1: D. Bauer and F. Ceccherini, "A numerical ab initio study ..."}

\pagebreak

\thispagestyle{empty}

\unitlength1cm
\begin{picture}(14,15)
\put(0,0){\bild{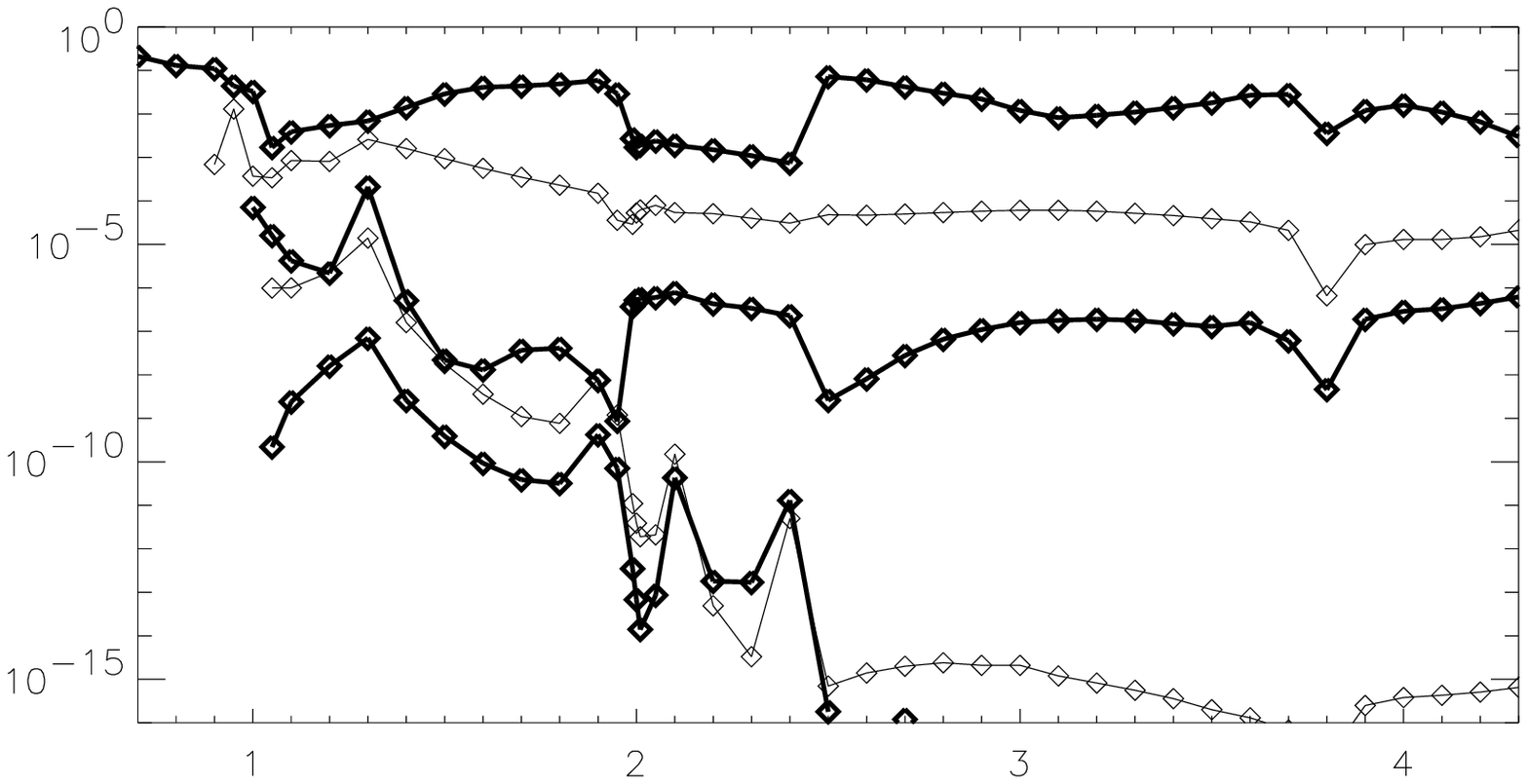}{14cm}}
\put(0.,7.4){\bild{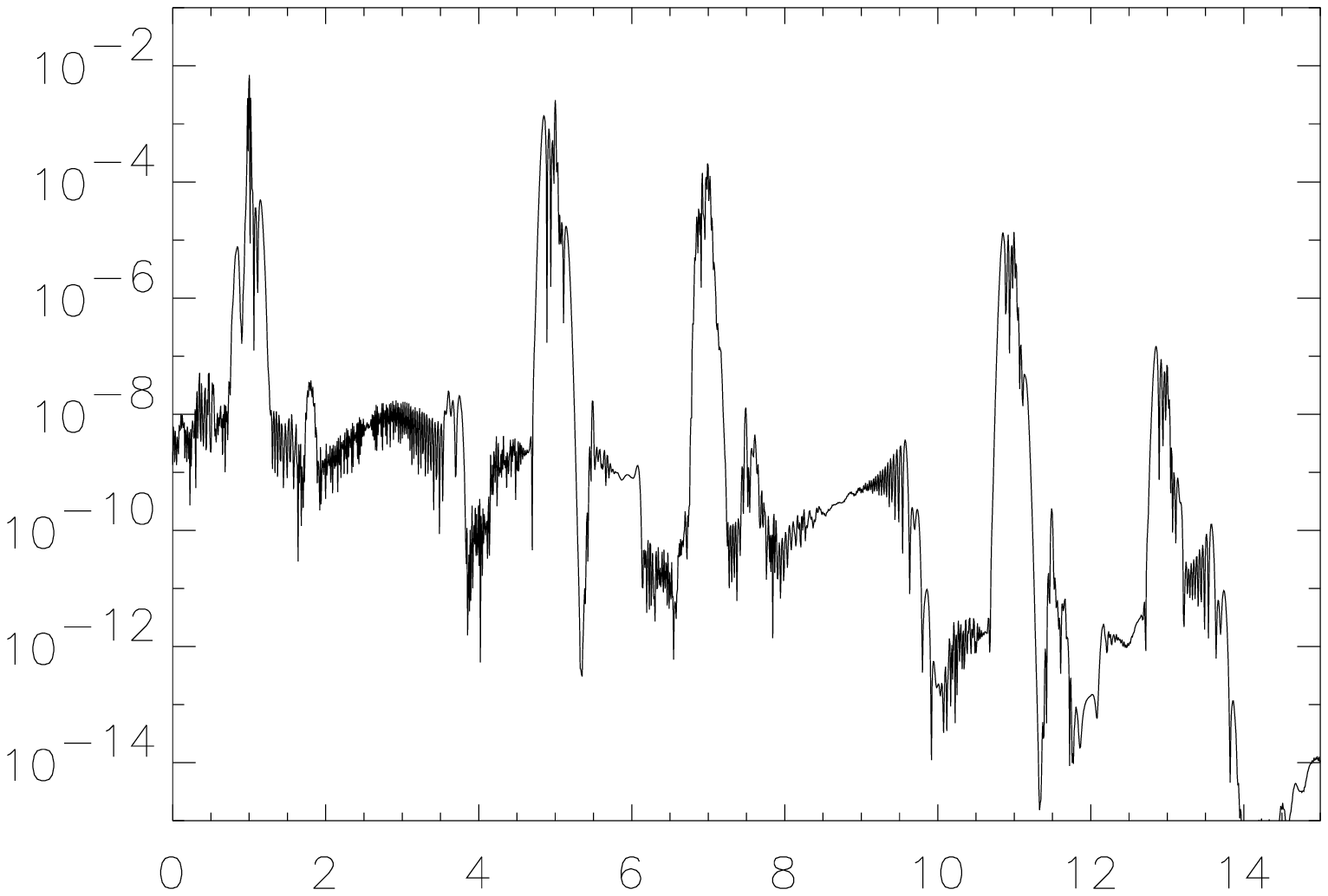}{4cm}}
\put(5,7.4){\bild{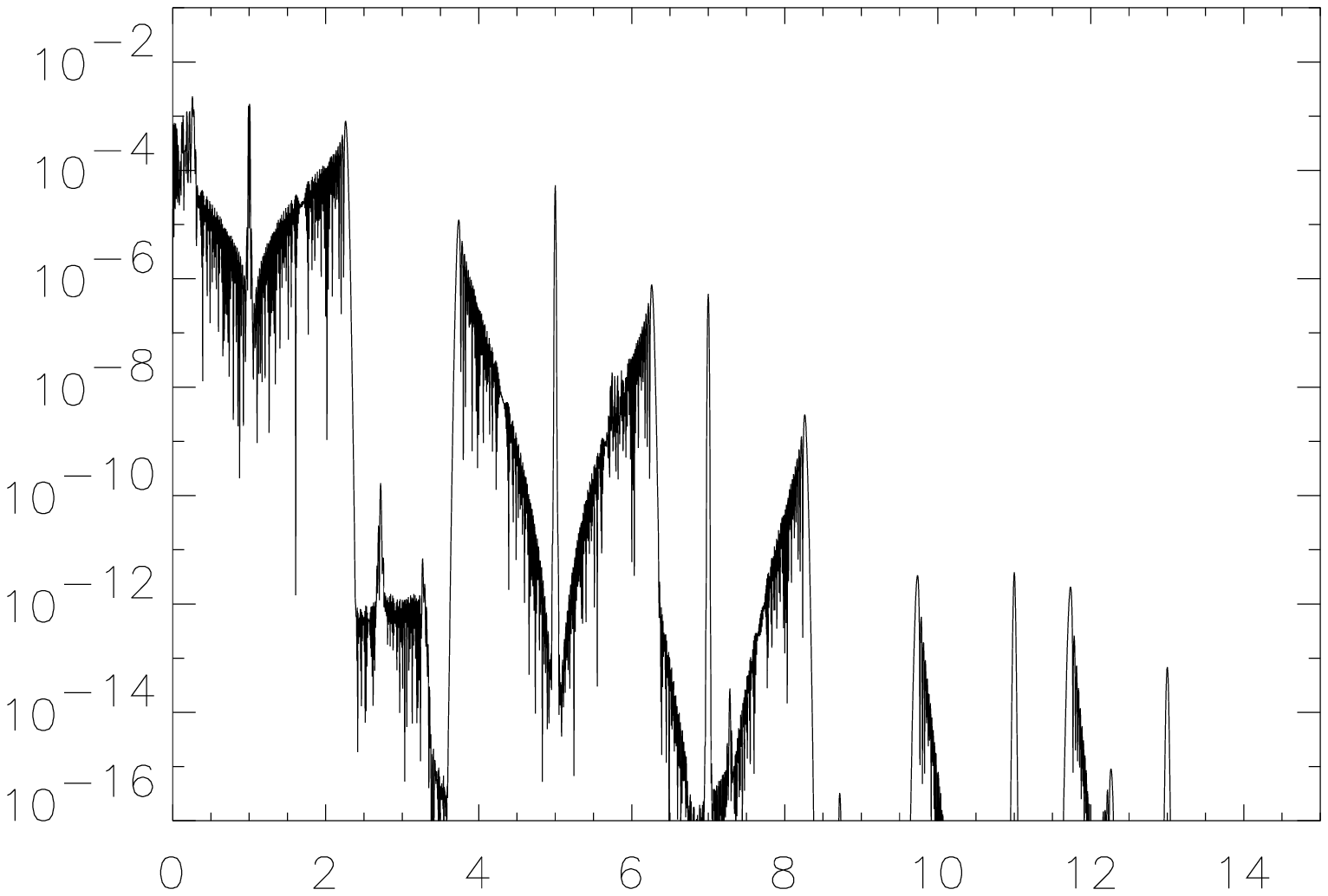}{4cm}}
\put(9.5,7.4){\bild{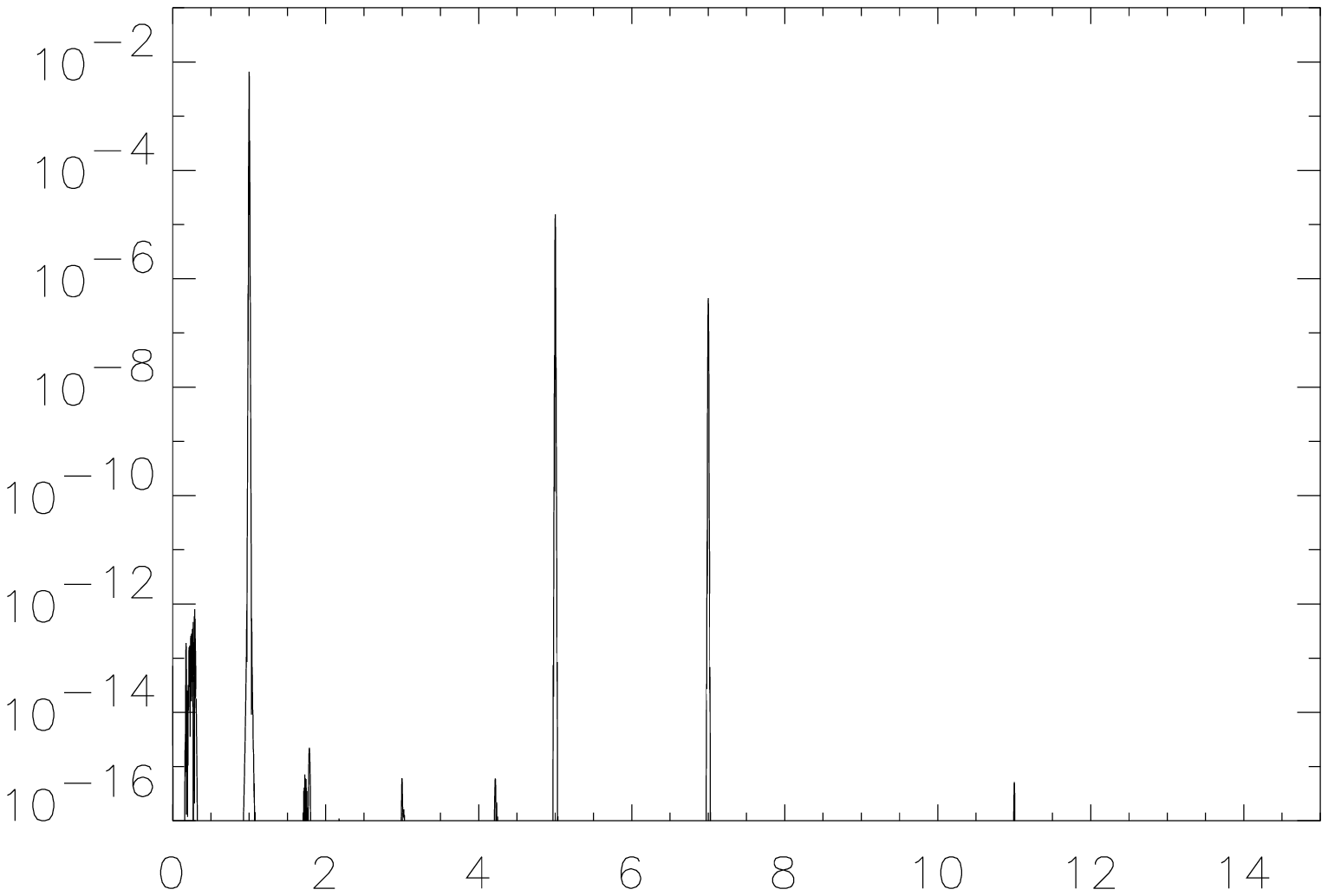}{4cm}}
\put(2.23,7.6){\vector(2,-1){1.2}}
\put(11.75,7.6){\vector(2,-1){1.2}}
\put(6.85,7.6){\vector(-2,-1){1.2}}
\put(9,6.5){1}
\put(9,5.65){5}
\put(9,4.7){7}
\put(8.7,1.8){11}
\put(2,3){13}
\put(7,0){$\omega/\Omega$}
\put(-0.3,2){\begin{sideways}Harmonic yield (arb.\ u.)\end{sideways}}
\end{picture}

\vfill

{\bf Fig. 2: D. Bauer and F. Ceccherini, "A numerical ab initio study ..."}

\pagebreak

\thispagestyle{empty}

\unitlength1cm
\begin{picture}(14,15)
\put(0,0){\bild{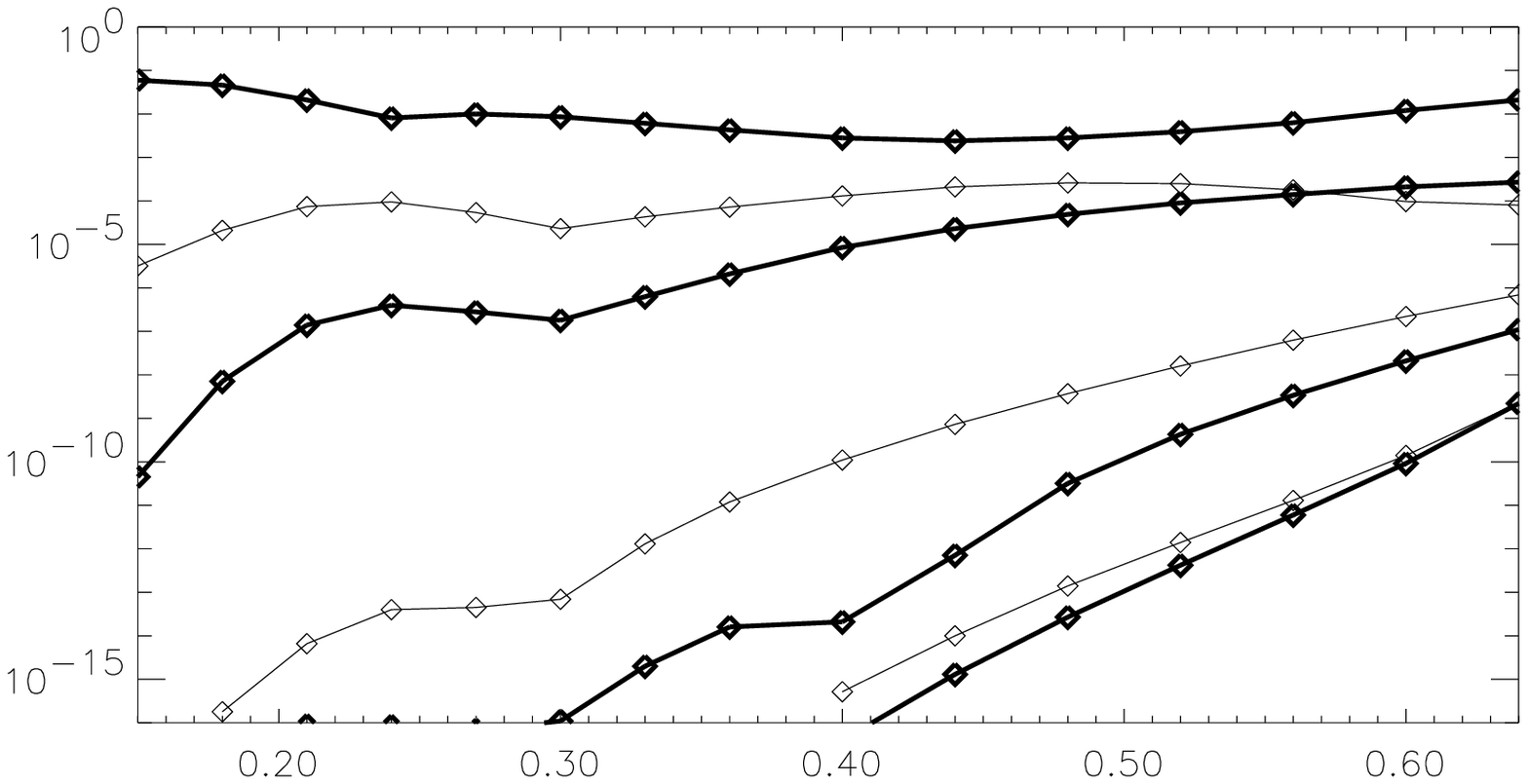}{14cm}}
\put(0.,7.4){\bild{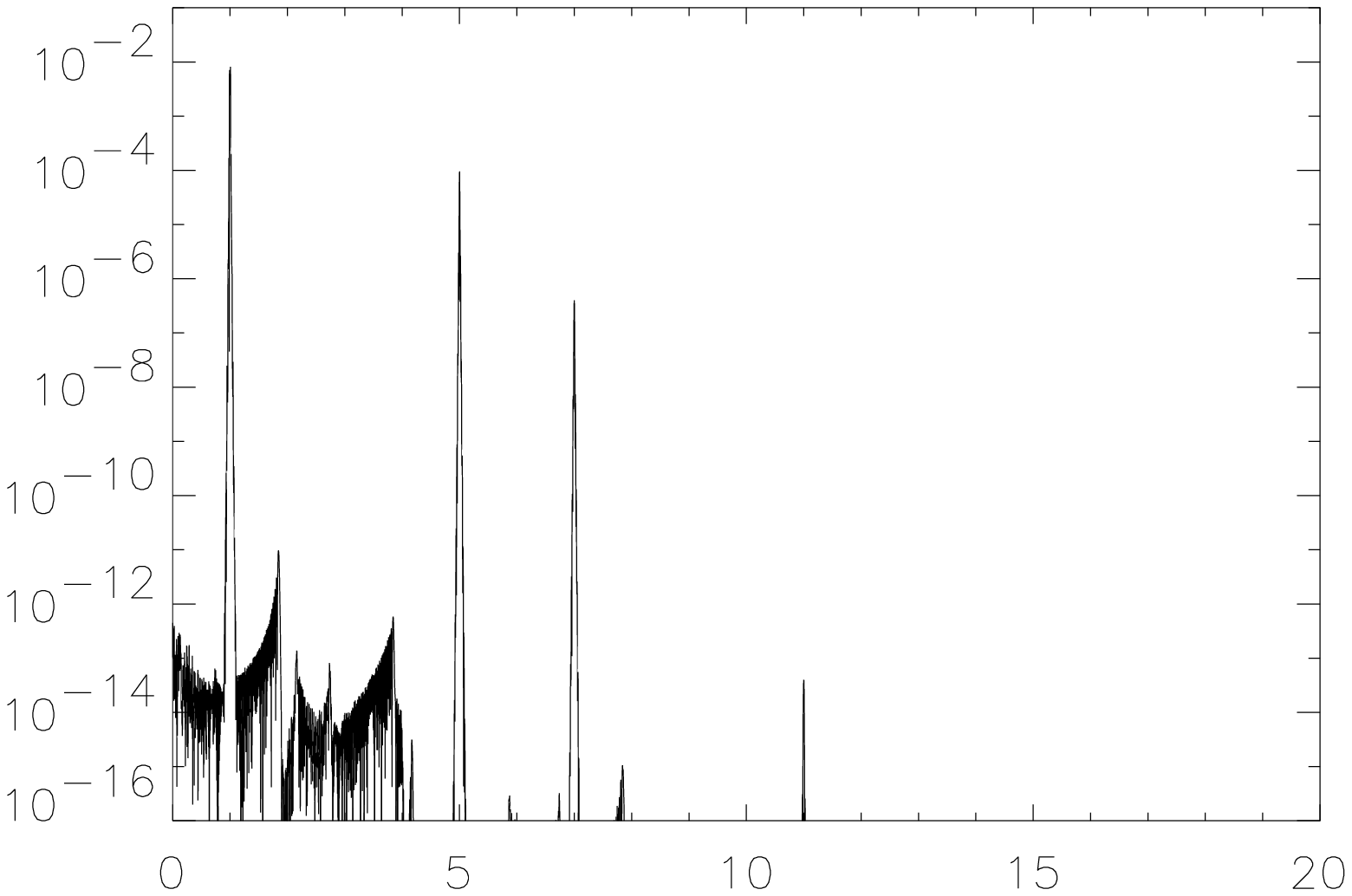}{4cm}}
\put(4.7,7.4){\bild{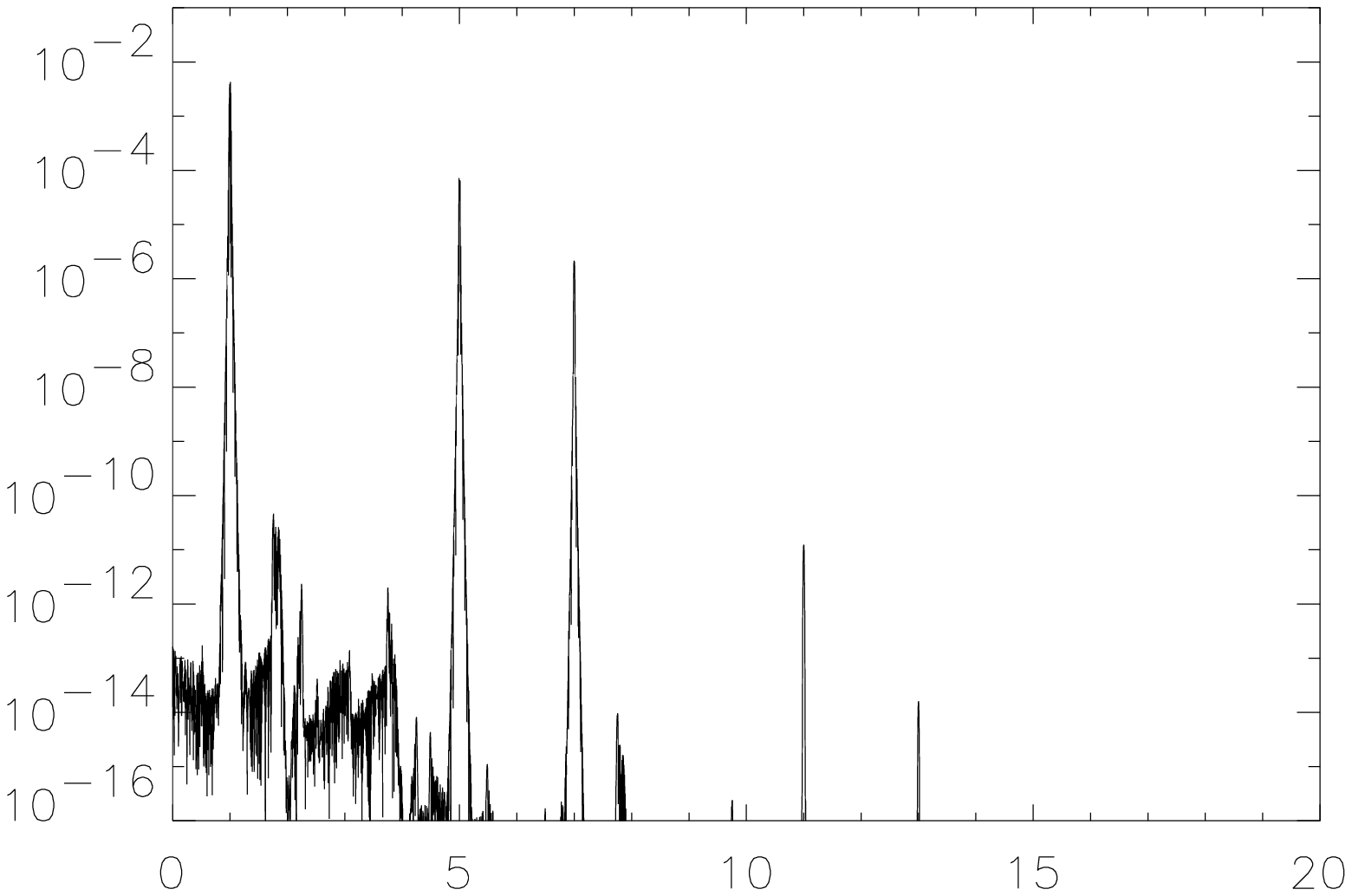}{4cm}}
\put(9.5,7.4){\bild{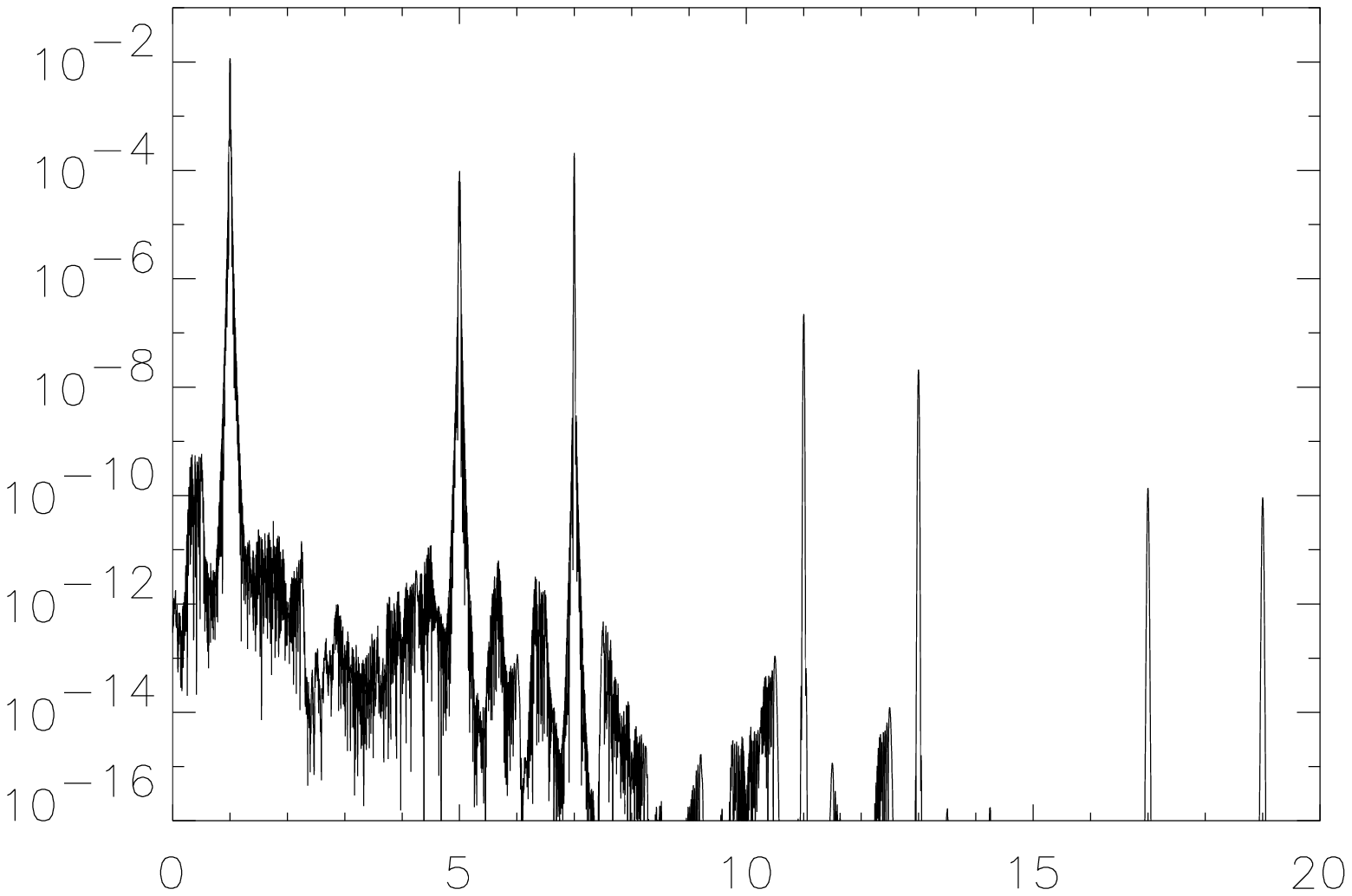}{4cm}}
\put(2.36,7.6){\vector(2,-1){1.2}}
\put(11.7,7.6){\vector(1,-1){0.6}}
\put(7.1,7.6){\vector(-1,-1){0.6}}
\put(3,6.5){1}
\put(3,5.65){5}
\put(3,4.7){7}
\put(8.7,3.8){11}
\put(8.8,3){13}
\put(10,2.7){17}
\put(9,1.3){19}
\put(6.5,0){$\Edach$ [a.u.]}
\put(-0.3,2){\begin{sideways}Harmonic yield (arb.\ u.)\end{sideways}}
\end{picture}

\vfill

{\bf Fig. 3: D. Bauer and F. Ceccherini, "A numerical ab initio study ..."}

\end{document}